\documentclass[aps,pra,floatfix,nofootinbib,showpacs,twocolumn]{revtex4-1}
\usepackage{lipsum}
\usepackage{amsmath}
\usepackage{epsfig}
\usepackage{color}
\usepackage{endnotes}
\usepackage{textcomp}

\newcommand{\be}{\begin{equation}}
\newcommand{\ee}{\end{equation}}

\newcommand{\ber}{\begin{eqnarray}}
\newcommand{\eer}{\end{eqnarray}}

\begin{document}
%\title{ A Translation-Invariant Density Functional }
% \title{ A Many-Body Density Functional }
  \title{ A Many-Body Density Energy Functional }

\author{A. Kievsky$^1$, G. Orlandini$^{2,3}$, M. Gattobigio$^4$}

\affiliation{$^1$Istituto Nazionale di Fisica Nucleare, Largo Pontecorvo 3, I-56100 Pisa, Italy}
\affiliation{$^2$Department of Physics, University of Trento, I-38123 Trento, Italy}
\affiliation{$^3$INFN-TIFPA Trento Istitute for Fundamental Physics and Applications,
      Via Sommarive 14, 38123 Trento, Italy}
\affiliation{$^4$ Universit\'e C\^ote d'Azur, CNRS, Institut  de  Physique  de  Nice,
1361 Route des Lucioles, 06560 Valbonne, France }

\begin{abstract}
The Hohenberg-Kohn theorem and the Kohn-Sham equations, which are at the basis of the Density Functional 
Theory, are reformulated in terms of a particular many-body density, 
which is translational invariant and therefore is relevant for self-bound systems. 
In a similar way that there is a unique relation between the one-body  density and the external 
potential that gives rise to it, we demonstrate that there is a unique relation between 
that particular many-body density and a definite many-body potential.
The energy is then a functional of this  density and its
minimization  leads to 
the ground-state energy of the system. As a proof of principle, the analogous of the Kohn-Sham
equation is solved in the specific case of $^4$He atomic clusters,
to put in evidence the advantages of this new formulation in terms of physical insights.

\end{abstract}

\bigskip

%keywords: 
\maketitle
\bigskip

\section{INTRODUCTION}

The essential idea behind the Density Functional Theory (DFT) is to reduce the complexity of the 
solution of the many-body Schr\"odinger equation to a much
tractable problem given in terms of the one-body density,  avoiding in this way the explicit
reference to the many-body wave function. In fact, Hohenberg and Kohn (HK)
demonstrated that the ground-state energy of a quantum system is
a functional of the one-body density $n(\vec r)$ and could, in principle,  
be obtained from a minimization procedure~\cite{hohemberg1964}. 
Moreover, Kohn and Sham (KS) demonstrated that the one-body
density can be represented by a non-interacting  system placed inside a particular  external field~\cite{kohn1965}.

The success of DFT for various many-body systems, for instance atoms, molecules, 
and the condensed phases has been enormous and its formulation and details are currently
part of books in which many of the recent applications are discussed (see for example
the books by G.F. Giuliani and G. Vignale~\cite{vignale}, by E. Lipparini~\cite{lipparini},
and by R.G. Parr and W. Yang~\cite{parr} and references therein). 
In the last fifteen years also 
nuclear physicists have devoted attention to it and carried out a considerable activity, 
starting from the seminal works of Refs.~\cite{bender2003,schunck}. 
Since nuclei are self-bound systems the original formulation of the HK theorem, which starts 
from a Hamiltonian containing a single particle external field, cannot be applied in a 
straightforward way. Such an external field, in fact, breaks the translation/Galileian invariance 
required by a self-bound system. To this problem a series of works have been devoted proposing 
different 
solutions~\cite{Engel2007,Giraudetal2008,Barnea2007,Giraud2008,Messud2009,Duguet2010,Chamel2010,Messud2013,marino}.

The aim of the present work is to suggest a formulation of the DFT that replaces the one-body density 
with a particular many-body density, and that, at the same time, fulfills the mentioned invariance.
More explicitly: the traditional energy functional $E[n(\vec r)]$ is replaced by $E [\nu(\rho)]$ 
where $\nu$ is a different density, expressed in terms of the so called hyperradius $\rho$, 
a collective variable depending on all interparticle distances
\begin{equation}
\rho^2= \frac{2}{N}\sum_{i<j} (\vec r_i - \vec r_j)^2 \, ,
\end{equation}
with $N$ the number of particles (of equal masses).
In a similar way that there is a unique relation between the one-body
ground-state density and the external potential, we demonstrate that there is a 
unique relation between $\nu(\rho)$ and a definite many-body potential $W(\rho)$.
The minimization of $E [\nu(\rho)]$ with respect to $\nu(\rho)$ leads to an equation whose solution allows to
know the ground-state energy of the system, if  the correct
information is included in $W(\rho)$. The search for the correct KS energy functional $E[n]$ is replaced by 
the search for the correct $E[\nu]$.
This approach lets us envisage the possibility that, in practical 
applications, the new functional $E [\nu(\rho)]$  might better take into account 
the complex many-body dynamics of a strongly interacting self bound system.

In Section~\ref{FORM} we introduce the new variable $\rho$ within the general so called Hyperspherical Harmonics 
(HH) formalism. In Section~\ref{DF} we define the associated density $\nu(\rho)$, the energy functional 
$E [\nu(\rho)]$  and the analogous of the KS approach, demonstrating a one to one relation 
between $\nu(\rho)$ and a many-body potential  $W(\rho)$. 
%The form of this potential is discussed in section~\ref{W}.
As a practical application, in Section~\ref{bosons} a particular energy functional is suggested for the much 
studied  bosonic self-bound systems, namely $^4$He clusters. Surprisingly, satisfactory results are obtained.
Further considerations and outlooks are found in Section~\ref{outlook}.

\section{FORMALISM}\label{FORM}

A convenient set of  translation-invariant coordinates useful to describe an interacting $N$-body system 
are the $N-1$ Jacobi vectors  $\vec \xi_i, i=1,\ldots N-1$ defined  (for equal masses $m$), as
\begin{equation}
\vec \xi_{N-j}=\sqrt{\frac{2j}{j+1}} (\vec r_{j+1}-\vec R_j), \,\,\, j=1,\ldots,N-1  
\end{equation}
with $\vec R_j=(1/j)\sum_{i=1}^j \vec r_i$. The $(3N-3)$ independent Jacobi coordinates can then be transformed 
into a set of as many independent {\it hyperspherical coordinates} (HC) \cite{Efros1972} 
consisting in the {\it hyperradius} $\rho$ 
\begin{equation}
 \rho = \sqrt{\sum_{i=1}^{N-1} \xi_i^2}\,,
\end{equation}
and $(3N-4)$ angles
$\Omega=(\hat \xi_1, \ldots, \hat \xi_{N-1}, \phi_2,\ldots,\phi_{N-1})$, with the {\it hyperangles} $\phi_i$ defined by
\begin{equation}
\cos\phi_i=\frac{|\vec\xi_i|}{\sqrt{\xi_1^2+\ldots +\xi_i^2}}\, ,\,\,\, i=2,\ldots,N-1 \, .
\end{equation}
Notice that the hyperradius $\rho$ is a (translation-invariant) many-body variable,  
involving  all particle distances. In fact it can be shown that  
\begin{equation} 
\rho^2 = \frac{2}{N}\sum_{i<j}^N 
 (\vec r_i - \vec r_j)^2=2\sum_i^N (\vec r_i - \vec R_N)^2 \, ,
\label{eq:radii}
\end{equation}
with $\vec R_N$ the center of mass position.
%In particular, the moduli of the Jacobi coordinates can be written in terms of $\rho$ and the
%hyperangles~\cite{gattobigio2009} and 
In terms of the hyperspherical coordinates the volume element $dV^{3N-3}=d\vec \xi_1\ldots d\vec \xi_{N-1}$ becomes
\begin{equation}
 dV^{3N-3}= \rho^{3N-4}d\rho\, d\Omega \, 
\end{equation}
(for notation and further details see e.g. Refs.~\cite{fabre1983,Barnea2000,gattobigio2009}).

Let us consider the typical translation-invariant Hamiltonian of a system characterized by mutual interactions
\begin{equation}\label{H}
 H= T   + \sum_{i<j} V_{ij} + \sum_{i<j<k}V_{ijk} + ...\equiv T+ V( \rho, \Omega) \, ,
\end{equation}
where $T$, the kinetic energy,
expressed in terms of the HC, $ \rho$ and $\Omega$, assumes a rather familiar form
 \begin{equation}
 T  =-\frac{\hbar^2}{2m}\left(\frac{\partial^2}{\partial \rho^2} + \frac{3N-4}{ \rho}\frac{\partial}{\partial \rho}
 + \frac{\hat K^2(\Omega)}{ \rho^2}\right)\,.
\end{equation}
%One will notice the analogy with the two-body kinetic energy expressed in polar coordinates. 
The operator $\hat K$, called the hyperangular momentum, has a complete set 
of orthonormal eigenfunctions ${\cal Y}_{[K]}(\Omega)$ called  hyperspherical harmonics, 
  that satisfy the following eigenvalue equation
\begin{equation}
\hat K^2 {\cal Y}_{[K]}(\Omega)= K (K+3N-5) {\cal Y}_{[K]}(\Omega)\,,
\end{equation}
where $K$ is called the grand angular quantum number and $[K]$ indicate all the other relative $3 N-4$ 
good quantum numbers.
% identifying the order of the homogeneous polynomial $\rho^K {\cal Y}_{[K]}(\Omega)$

The $N$-body wave function, $\Psi( \rho,\Omega)$, can in principle be expanded in terms of the 
HH functions up to some $[K_M]$ (in principle infinite) as
%\begin{eqnarray}\label{expansion}
%\Psi(\rho,\Omega)&=& \sum_{[K]=[K_m]}^{[K_M]} {\cal R}_{[K]}(\rho){\cal Y}_{[K]}(\Omega_N)\\
%&\equiv&\sum_{[K]=[K_m]}^{[K_M]} \left(\frac{u_{[K]}(\rho)}{\rho^{(3N-4)/2}}\right){\cal Y}_{[K]}(\Omega_N)\,, 
%\end{eqnarray}
\begin{equation}
\label{expansion}
\Psi(\rho,\Omega)= \rho^{-(3N-4)/2}
\sum_{[K]=[K_m]}^{[K_M]} {u_{[K]}(\rho)}{\cal Y}_{[K]}(\Omega_N)\,, 
\end{equation}
where $[K_m]$ is the set of quantum numbers consistent with the minimal value of the
grand angular quantum number $K_m$,  compatible with the permutational symmetry requirements.
In the case of spin-$0$ bosons $K_m=0$ (and so are all other good quantum number in $[0]$); 
for higher values of the spin or for fermions $K_m$ could be different from zero~\cite{kievsky2008}. 
%Moreover the HH set has to be extended to consider the corresponding
%internal degrees of freedom (see for example Ref.~\cite{kievsky2008}). 
%The set $[K_M]$ indicates the grand angular maximum value considered to achieve the desired 
%accuracy in the description of the system.

For normalizable wave functions one can then define the density 
\begin{equation}\label{rhodensity}
 \nu( \rho) \equiv \int d\Omega\,|\Psi( \rho,\Omega)|^2  = \rho^{-(3N-4)}\sum_{[K]=[K_m]}^{[K_{M}]} u^2_{[K]}(\rho)  ,
\end{equation}
normalized  as 
\begin{equation}
 \int_0^\infty d \rho \, \rho^{3N-4}\, \nu( \rho) 
= \int_0^\infty d\rho\,\left[\sum_{[K]=[K_m]}^{[K_M]}u^2_{[K]}(\rho)\right]  = 1 \, . 
\end{equation}
It is in terms of this density that, in the next section, the HK theorem will be reformulated.
%,however, following Levy's proof.
 
\section{The energy functional}\label{DF}

%Following Levy's proof of the HK theorem~\cite{Levy}, in the following we will prove a similar 
%theorem, where, however, the functional is  $E[\nu( \rho)]$ and refers to the Hamiltonian $H$ 
%in  Eq.~(\ref{H}), namely does not contain an external potential.

We start from the Hamiltonian in Eq.(\ref{H})
%, which for simplicity we write as  $H=T+V$, 
and define 
the functional $E[\nu]$ as the minimum of the energies obtained with all wave functions $\Psi$ that 
have the same $\nu( \rho)$: 
\begin{equation}\label{min}
% E[\nu]=min_{\Psi\rightarrow\nu}\langle \Psi|T+V|\Psi \rangle
 E[\nu]=\underset{\Psi\rightarrow\nu}{\text{min}} \langle \Psi|T+V|\Psi \rangle \, .
\end{equation}
Since $\nu( \rho)$ is an integral property of 
%the wave function 
$|\Psi \rangle $, there will be in principle 
an infinite number of normalizable functions having the same $\nu$. The functional $ E[\nu]$ 
is then defined as the minimum produced by all such functions.

Calling $E_0$ the ground state energy of $H$, and $|\Psi_0\rangle$ the corresponding wave function, 
the Rayleigh-Ritz variational principle establishes that
\begin{equation}
E[\nu]\geq E_0 \, ,
\end{equation}
and
\begin{equation}
% E[\nu_0] =min_{\Psi\rightarrow\nu_0} \langle Psi|T+V|\Psi \rangle = E_0 \, ,
  E[\nu_0] =\underset{\Psi\rightarrow\nu_0}{\text{min}} \langle \Psi|T+V|\Psi \rangle = E_0 \, ,
\end{equation}
where we have denoted by $\nu_0$ the density corresponding to the ground state wave 
function $|\Psi_0\rangle$. This last statement follows from the fact that when the set of functions
are selected among those having the density $\nu_0$, the true wave function $|\Psi_0\rangle $ is
included in that set and the global minimum $E_0$ is reached. 
The above statements imply that   
\begin{equation}
 \left.\frac{\delta E[\nu]}{\delta\nu}\right|_{\nu=\nu_0}=0\,.
\end{equation}
Let us consider now the particular case of a system interacting through an hypercentral
potential $W(\rho)$
\begin{equation}
H_W= T + W(\rho)\,.  
\end{equation}
Analogously to the central potential case for two particles, the ground state wave function is simply
\begin{equation}\label{expansion0}
\Phi_0(\rho,\Omega)=\rho^{-(3N-4)/2}w_{[K_m]}(\rho){\cal Y}_{[K_m]}(\Omega)\,, 
\end{equation}
namely it includes only the lowest term of the expansion given in Eq.(\ref{expansion}). The hyperradial
function $w_{[K_m]}$ and the ground state energy ${\cal E}_0$ can be obtained by solving a one dimensional 
differential equation (hyperradial equation).

%\lipsum[1]
\begin{widetext}
\begin{equation}
\left[
-\frac{\hbar^2}{m}
\left(
\frac{\partial^2}{\partial\rho^2}-
\frac{(3N-4)(3N-6)}{4\rho^2}+
\frac{K_m(K_m+3N-5)}{\rho^2}
\right) + W(\rho) 
-{\cal E}_0 \right] w_{[K_m]}(\rho)=0 \, ,
\label{eq:u0r}
\end{equation}
\end{widetext}
%\lipsum[1]
and the ground state density defined in Eq.~(\ref{rhodensity}) is
\begin{equation}
 \nu_0^W( \rho)  = \rho^{-(3N-4)} w^2_{[K_m]}( \rho )  
\end{equation}
normalized as 
\begin{equation}
 \int_0^\infty d \rho \, \rho^{3N-4}\,\nu_0^W( \rho) = 
 \int_0^\infty d \rho \, w^2_{[K_m]}( \rho)=1   \, .
\end{equation} 
On the other hand, after having defined 
$T_{min}[\nu]\equiv \underset{\Psi\rightarrow \nu}{\text{min}} \langle\Psi|T |\Psi\rangle$ 
and the density functional
\begin{eqnarray}\nonumber
E_W[\nu]&\equiv& \underset{\Psi\rightarrow \nu}{\text{min}} \langle\Psi|T+W|\Psi\rangle\\ 
&=& T_{min}[\nu] + \int d\rho \, \rho^{3N-4} \nu(\rho) W(\rho)\,,
\end{eqnarray} 
 ${\cal E}_0$ could as well be found by a minimization procedure, in fact
\begin{widetext}
\begin{equation}
\frac{\delta E_W[\nu]}{\delta  \nu}= \frac{\delta T_{min}[\nu]}{\delta  \nu}+\rho^{3N-4}W(\rho) =0
\Longleftrightarrow \nu=\nu_0^W\,\,\,\, {\rm and\,\,\,}{\cal E}_0 =E_W(\nu_0^W) \label{defW}
\end{equation}
\end{widetext}
Turning back to the energy functional of Eq.(\ref{min}) we impose the following requirement
\begin{equation}
 E[\nu]=E_W[\nu]
\label{eq:funn}
\end{equation}
namely
\begin{eqnarray}\label{crucial} \nonumber
 \underset{\Psi\rightarrow \nu}{\text{min}} \langle\Psi| T&+&V|\Psi\rangle 
= \underset{\Psi\rightarrow  \nu}{\text{min}} \langle\Psi| T+W|\Psi \rangle\\
 &=& T_{min}[\nu] + \int d\rho \,\rho^{3N-4} \nu(\rho) W(\rho) \,.
 %\,\,\,\,\,{\rm with}\,\,\,\,\, \nu_0( \rho)= \nu_{hc}( \rho),
\end{eqnarray}
The energy functional $E[\nu]$ is now represented by a system of particles interacting through this particular 
potential $W(\rho)$. Since for the ground state of $H=T+V$  one has $\frac{dE[\nu]}{d\nu}=0$, 
Eq.~(\ref{crucial}) formally defines the hypercentral potential $W(\rho)$  as 
%Formally the hypercentral potential is defined by Eq.~({\ref{defW})
\begin{equation}\label{fmin}
W(\rho)= -\frac{1}{\rho^{3N-4}} \frac{\delta T_{min}[\nu]}{\delta  \nu} \, .
\end{equation}
%Formally this defines a unitary transformation $U$ such that $(T+W)=U^{-1}(T+V)U$ and $|\Psi\rangle=U |\Phi\rangle$. This allows to interpret $W(\rho)$ as an effective potential. Moreover, 

The core of requirement (\ref{crucial})  is  that $\,\,W(\rho)$ gives the same density $\nu(\rho)$ as $V(\rho,\Omega)$. 
One can show that such a $W(\rho)$ is unique.\footnote{Here the argument is similar to that of Kohn-Sham one-body potential, namely, once the existence of $W$ is assumed, one can show its uniqueness.} 
The proof goes via a {\it  reductio ad absurdum} procedure. One assumes  that two hypercentral
potentials, $W_1( \rho)$ and $W_2( \rho)$ differing by more than a constant, exist in such 
a way that the two hamiltonians $H^W_1=T+W_1( \rho)$ and $H^W_2=T+W_2(\rho)$ have the same 
$\nu( \rho)$. Let us call $|\Phi_1\rangle$ and $|\Phi_2\rangle$ the respective
wave functions and ${\cal E}_1$ and ${\cal E}_2$ the corresponding energies. 
From the Rayleigh-Ritz variational principle the following condition holds
\begin{eqnarray}\nonumber
{\cal E}_1&<&\langle\Phi_2|H^W_1|\Phi_2\rangle = \\
&&\langle\Phi_2|H^W_2|\Phi_2\rangle +\langle \Phi_2|H^W_1-H^W_2|\Phi_2\rangle \\
{\cal E}_1&<& {\cal E}_2+ \int d \rho\,\rho^{3(N-4)}\, [W_1( \rho)-W_2( \rho)]\,\nu( \rho)    
\end{eqnarray}
The same can be repeated starting from ${\cal E}_2$ arriving at
\begin{equation}
{\cal E}_2<{\cal E}_1 + \int d \rho\,\rho^{3(N-4)}[W_2( \rho) - W_1( \rho)]\,\nu( \rho)   
\end{equation}
Summing both inequalities we arrive at the following contradiction ${\cal E}_1+{\cal E}_2< {\cal E}_1+{\cal E}_2$,
proving that the first assumption was wrong. 
Accordingly, it is proven that the density $\nu( \rho)$ uniquely determines the hyperradial potential
$W(\rho)$ that generates it.
Notice, by the way, that the same conclusion holds if a further generic ${\cal W}(\rho, \Omega)$ 
interaction is included in $H_1$ and $H_2$. 

The important conclusion is then that $E_0$ could be found either by  $\frac{\delta  E_W[\nu]}{\delta  \nu}=0$ 
or simply by solving Eq.~(\ref{eq:u0r}). 
In the traditional KS case the problem  is to guess the correct functional, here it is to guess the correct $W(\rho)$.
 
Eq.~(\ref{eq:u0r}) is the basic equation of the translational-invariant density functional theory discussed here. 
This equation has been obtained previously in the literature (see for example
Refs.\cite{fabre1979,fabre1982,greene1998}), however, 
in a different context, namely as the result of the lowest order HH expansion of the ground state 
wave function. In our case,  in view of the  unique relation between $W(\rho)$ and  the density $\nu(\rho)$, 
this equation provides the way to obtain the right energy functional, and therefore the right ground state energy, 
for any number of particles. 
 
As a first application, in the next section the functional $E[\nu]$ is constructed in the case of atomic
clusters of bosonic helium. The case of fermions is postponed to a forthcoming work.

\section{Application to Atomic clusters}\label{bosons}

We consider clusters of atomic $^4$He, largely discussed in the literature.
Helium drops and the homogeneous system have been extensively studied using realistic 
He-He potentials. A rather successful one is the Aziz HFD-HE2 He-He 
potential~\cite{aziz1}, which we will take as reference potential and, 
for the purpose here, its results are considered equivalent to experimental data.

Another interesting approach to helium clusters starts from the observation that the dimer 
of $^4$He has a binding energy 
of about $1\,$mK, three orders of magnitude less than the typical energy scale of 
$\hbar^2/m r_{vdW}^2=1.677\,$K, with $\hbar^2/m=43.281307\,$K$\, a_0^2$ and $r_{vdW}=5.08\,a_0$ 
the corresponding van der Waals length.
Moreover, the two-body scattering length has been estimated to be $a\approx 190\,a_0$, twenty times
larger than $r_{vdW}$. 
In the limiting case, $a\rightarrow\infty$, the system is located at the unitary limit, well
suited for an effective expansion of the interaction. 
In the spirit of an effective field theory devoted to describe system with large values
of the two-body scattering length~\cite{bira,bazak},
the first term of this expansion is a contact interaction between the two helium atoms.
However, as it is well known, the three-body system (as well as larger systems) collapses, even if
the contact interaction is set to produce an infinitesimal binding energy. This phenomenon
is known as the Thomas collapse~\cite{thomas} and it is remedied by the introduction of a (contact) three-body force
set to correctly describe the trimer energy $\epsilon_3$. Accordingly, the leading order (LO) of this
effective theory has two terms, a two-body term and a three-body term, associated with two constants,
named low-energy constants (LECs), needed to determine their strengths, usually fixed by $\epsilon_2$ and $\epsilon_3$.

For the only purpose of determining $W(\rho)$, and inspired by the effective theory just described above,  
we introduce the following two- and three-body potentials
 \begin{equation}\label{V2V3}
 \! V^{[2]}_{LO} = \sum_{i<j}A e^{-r_{ij}^2/\alpha^2}\,,\,\,\,\,\,
  V^{[3]}_{LO} = \sum_{i<j<k}
 B e^{-r_{ijk}^2 /\beta^2}\,,
\end{equation}
where $r_{ij}^2\equiv (\vec r_i-\vec r_j)^2$ and 
$r_{ijk}^2=(\vec r_i -\vec r_j)^2+(\vec r_i -\vec r_k)^2 + (\vec r_j -\vec r_k)^2$. 
the natural choice is to consider $W(\rho)$ as a sort of mean hypercentral field and obtain 
it by  averaging on the hyperangular part of the ground state wave function.
\begin{equation}
W_A(\rho) =A\,\frac{N(N-1)}{2} \int d\Omega e^{-r_{12}^2/\alpha^2} 
|{\cal Y}_{[0]}(\Omega)|^2\,,  
\end{equation}
\begin{equation}
W_B(\rho) = B\,\frac{N(N-1)(N-2)}{6} \int d\Omega e^{-r _{123}^2/\beta^2} 
|{\cal Y}_{[0]}(\Omega)|^2  
\end{equation}
%(in the case of $^4$He clusters the minimal value of $[K_m]=0$).  
(for spin 0 systems the minimal value of $[K_m]=0$).  
Performing the integrals one has   
\begin{eqnarray}\nonumber
&&W(\rho)=A\,\frac{N(N-1)}{2} \,
{\cal M}\left(\frac{3}{2},\frac{3N-3}{2},-\frac{\rho^2}{\alpha^2}\right)+\\&+& B\,
 \frac{N(N-1)(N-2)}{6}  \,
{\cal M}\left(3,\frac{3N-3}{2},-\frac{3\rho^2}{\beta^2}\right)\,,
\label{eq:uhc1}
\end{eqnarray}
where the function  ${\cal M}(a,b,c)$ is a confluent hypergeometric function~\cite{timofeyuk1,timofeyuk2}.

Having found an expression for $W(\rho)$ one can now give a prediction for the ground 
state energy of clusters of any number $N$ of bosons by solving the simple equation.
\begin{equation}
\left[-\frac{\hbar^2}{m}\left(\frac{\partial^2}{\partial\rho^2}-
\frac{(3N-4)(3N-6)}{4\rho^2}\right)+ W(\rho)-{\cal E}_0\right] w_0(\rho)=0.
\label{eq:ubr}
\end{equation}

The four parameters in $W(\rho)$ give us the opportunity to relate 
the two functionals, $E_W[\nu]$ and $E[\nu]$ as follows.
The helium dimer represented by the Aziz potential
has a single bound state with energy $\epsilon_2=0.83012\,$mK, a 
scattering length $a=235.547\,a_0$ and an effective range $r_e=13.978\,a_0$. 
Fitting $\alpha$ and $A$  to the corresponding HFD-HE2 values the gaussian parameters
result $\alpha=10.0485\,a_0$ and $A=-1.208018\,$K. 
 
Several choices are possible to determine the other two parameters $(B,\beta)$. One could choose e.g. 
to fit the trimer and tetramer binding energies~\cite{artur1,artur2}. In view of the fact that $W(\rho)$ has to 
account for energies at any $N$, we think more expedient to obtain couples $(B,\beta)$ values, all fitting the 
tetramer binding energy. So we solve  
Eq.(\ref{eq:ubr}) for the four-body system and require ${\cal E}_0 =  0.5332\,$K, 
the HFD-HE2 value~\cite{pandha1}. 
We observe substantial independence from the three-body range
$\beta$ for the lowest $N$ values with the overall best description 
inside the interval $7.5\, a_0 < \beta < 9.0\, a_0$,
the central value is $\beta=8.33\, a_0$ and $B=7.211\,$K the corresponding strength. 
\begin{figure}[h]
\includegraphics[scale=0.33]{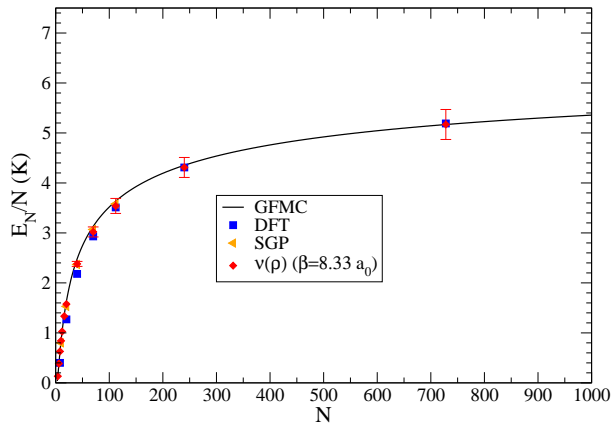}
\caption{
Binding energy per particle obtained solving Eq.(\ref{eq:ubr}) with $7.5\, a_0 < \beta < 9.0\, a_0$,
the (red) diamonds highlight the best case $( B,\beta)=(7.211\, K, 8.33\, a_0)$ with the spread
denoted by the error bars visible for $N\ge 40$.
The GFMC results for the HFD-HE2 potential (black solid line),
the DFT results of Ref.~\cite{dalfovo} (blue squares) and the results of the
soft gaussian potential (SGP) of Refs.~\cite{artur1,artur2} (orange triangles) are shown too.}
\label{fig:fig1}
\end{figure}
\begin{figure}[h]
\includegraphics[scale=0.33]{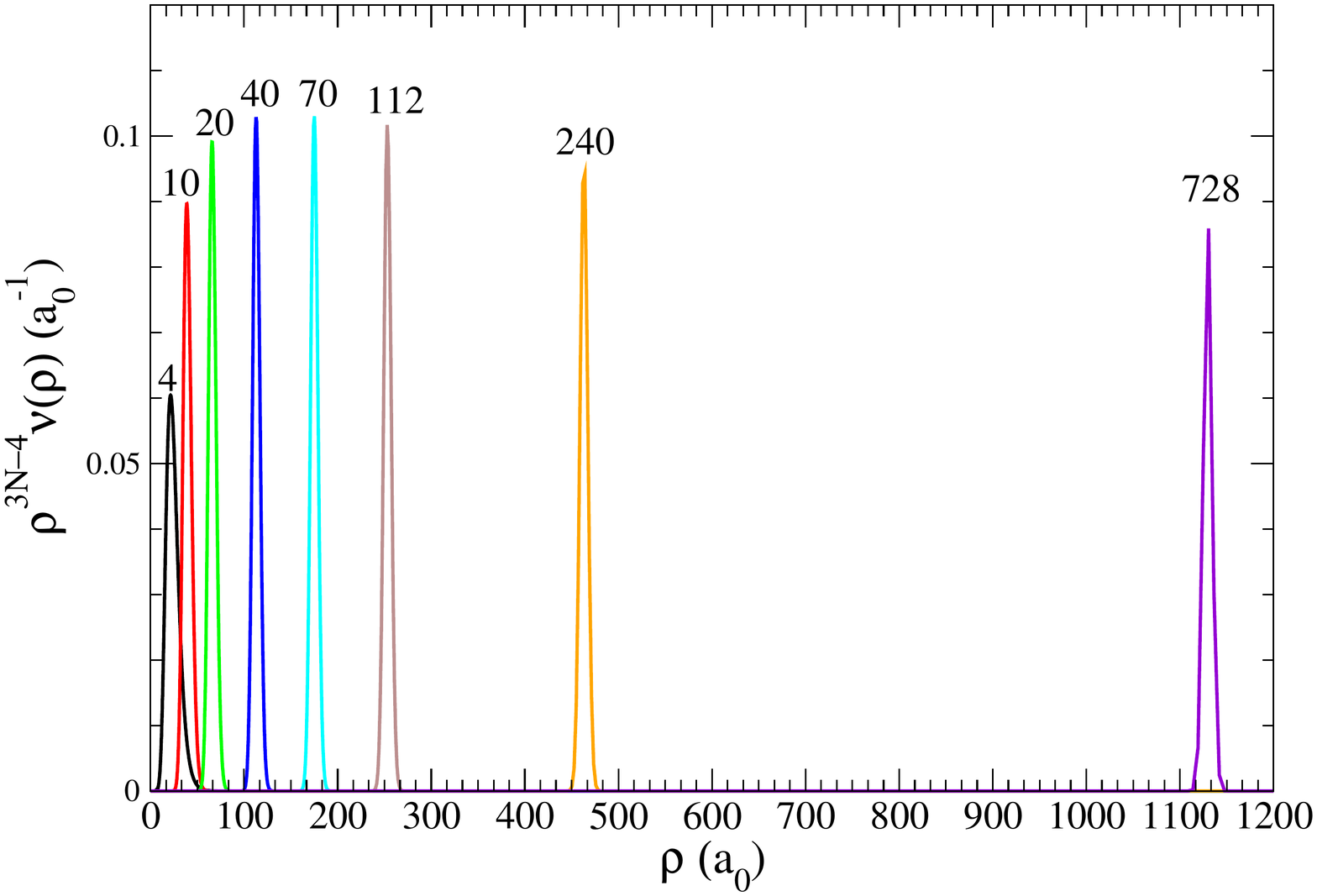}
\includegraphics[scale=0.33]{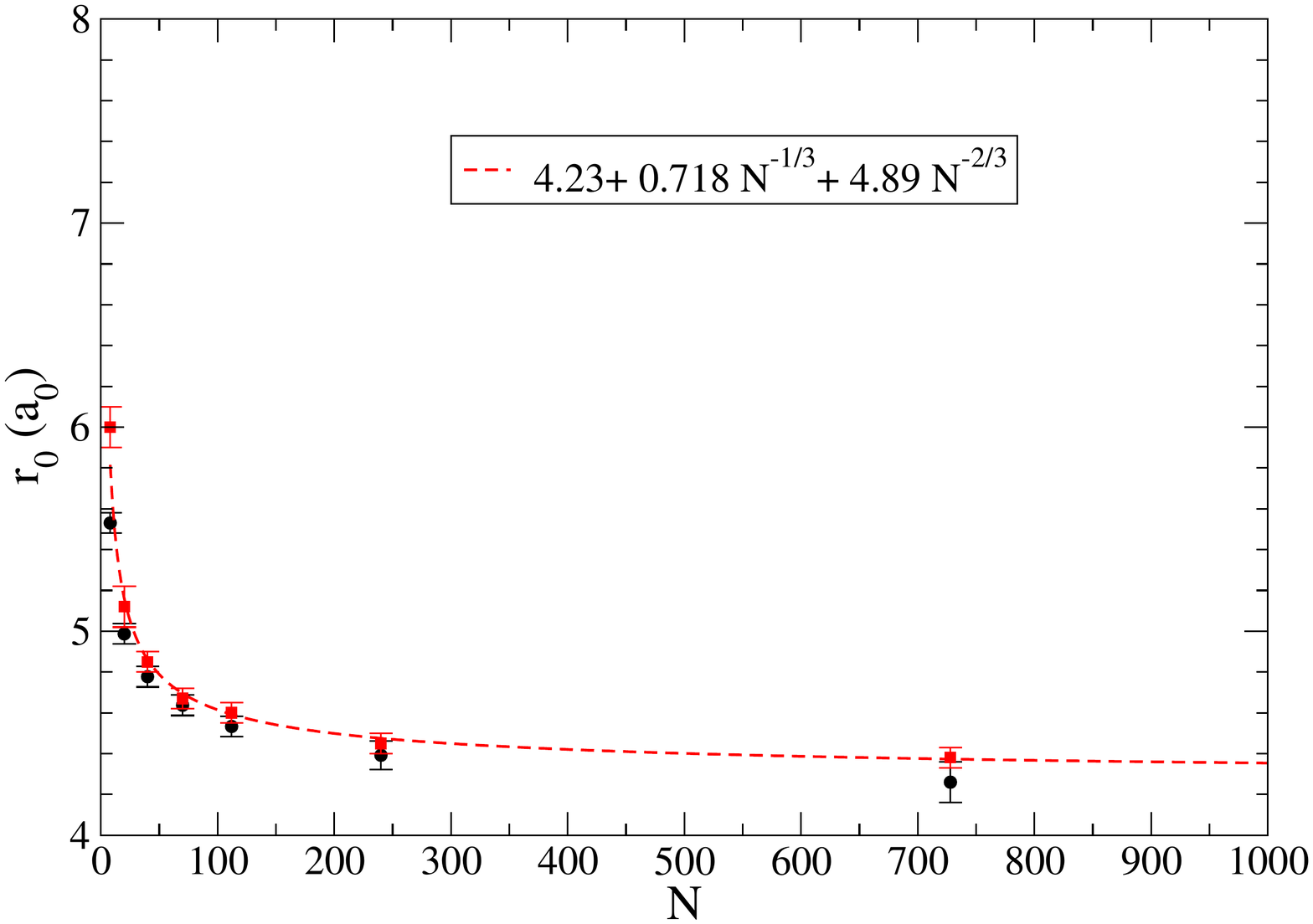}
\caption{ The (reduced) many-body density $\nu(\rho)$ for selected number of particles (upper panel).
The unit radius  $r_0$ (black solid points) (lower panel) with
error bars corresponding to variations of $\beta$ in the interval $7.5\, a_0 < \beta < 9.0\, a_0$. 
For the sake of comparison the GFMC results \cite{pandha1} are shown too (red solid points)
together with a fit to these values represented by the (red) dashed line in units of $a_0$. }
\label{fig:fig2}
\end{figure}
Having determined the parametrization of $W(\rho)$ we proceed to solve Eq.(\ref{eq:ubr}) for
increasing values of $N$ to determine the binding energy per atom $E_N/N$ as a function of the number 
of atoms $N$. The results are shown in Fig.~\ref{fig:fig1}. The red diamonds are the results
obtained with $( B,\beta)=(7.211\, K, 8.33\, a_0)$, the error bars show results with $\beta$ 
varying inside the interval $7.5\, a_0 < \beta < 9.0\, a_0$. For the sake of comparison we show
the Green Function Monte Carlo (GFMC) results of Ref.~\cite{pandha1} (black
solid line), the DFT results of Ref.~\cite{dalfovo} (blue diamond) and the results of the
soft gaussian potential (SGP) of Refs.~\cite{artur1,artur2} (orange triangles). 
Unexpectedly the four-parameter hyperradial potential $W(\rho)$ has
sufficient information to reproduce the $E_N/N$ behavior. 
%and to predict extremely
%well the experimental value of $7.1$ K of the homogeneous system. 
As it is evident in Fig.\ref{fig:fig1}, the results of Eq.(\ref{eq:ubr})
are of similar quality as those using more sophisticated potentials and methods,
giving support to the formalism presented. 

As we will see below, for large values of $N$ the solutions of Eq.(\ref{eq:ubr}) are very much localized, 
the kinetic energy
gets a small fraction of the total energy which tends to equal the minimum of $W(\rho)$. 
Using the asymptotic form of the confluent hypergeometric functions such a minimum results
\begin{equation}
W_m=-\frac{3^4}{8}\frac{A^2}{B}\left(\frac{\alpha}{\beta}\right)^6 N \, .
\end{equation}
The potential parameters selected to reproduce the binding energy per particle for low $N$ values
predict $E_N/N\rightarrow |W_m|/N=6.4\pm 0.7 \,$K to be compared to the experimental value of $7.1$ K 
of the homogeneous system. Though this is a remarkable result considering the minimal
information used to determine $W(\rho)$, the above relation gives a further condition 
that might be used in determining the potential parameters.

To conclude the analysis of the results, in Fig.2 (upper panel) we show the (reduced) 
many-body density for selected number of particles. As can be seen from the figure the 
density is extremely localized around a particular value of $\rho$, which increases almost
linearly with $N$. The behavior of $\nu(\rho)$ indicates a very compact object, not compressible;
in fact, lower values of $\rho$ allowing the particles to be closer are discouraged as well
as larger values, which would indicate possible clusterizations.

The many-body density $\nu(\rho)$ can be used to calculate the mean square radius. 
Defining $\vec r$ the position  of a generic particle with respect to the center of mass, 
using Eq.(\ref{eq:radii}) it results $\langle r^2 \rangle =\frac{1}{2 N } \langle \rho^2 \rangle $ with
\begin{equation}
 \langle \rho^2 \rangle = \int \rho^2 \nu(\rho) \rho^{3N-4} d\rho
\end{equation}
In Fig.2 (lower panel) the unit radius, $r_0(N)=\sqrt{5/3}\,\langle r^2\rangle^{1/2}/N^{1/3}$, 
(black solid points) is shown as a function of $N$ with error bars
corresponding to variations of $\beta$ in the interval indicated above and
it is compared to the GFMC results of Ref.\cite{pandha1} (red solid points). 
The agreement is evident, the $r_0$ values obtained from the best parametrizations of $W(\rho)$
reproduce the GFMC results better than $5\%$. Moreover, in the large $N$ limit, $W_m$ is located at
\begin{equation}
\rho_m=\left(\frac{2\cal B}{|{\cal A}|}\right) N^{5/6}
\end{equation}
with ${\cal A}=\frac{A}{2}(\frac{3}{2})^{3/2}\alpha^3$ and ${\cal B}=\frac{B}{6}(\frac{1}{2})^3\beta^6$.
The unit radius tends to $r_0\rightarrow\sqrt{5/6} (\frac{2\cal B}{|{\cal A}|})^{1/3}=4.1\pm 0.2 \,a_0$,
extremely close to the GFMC results at $N\rightarrow\infty$.

\section{Conclusions and outlook}\label{outlook}

In this work we formulate a density functional approach in terms of the
density $\nu(\rho)$. Such a 
density depends on the hyperradius $\rho$, a translation invariant variable of collective nature, 
because connected to the sum of the distances between the particles. 
It is shown that the functional $E[\nu]$ is governed by a unique (unknown) hyperradial potential $W(\rho)$. 
The solution of a single hyperradial equation with such an hyperradial potential allows to determine 
the binding energy for any number of particles in a straightforward way.

We have applied this framework to the bosonic case, focusing on $^4$He clusters. The guess for $W(\rho)$ has been
inspired by the effective theory approach together with a generalization of the mean field concept. Extremely
satisfying results have been found. The key point has been to use the range of the three-body interaction, $\beta$,
to fine tune the hyperradial potential $W(\rho)$. The extension to treat trapped systems is underway.
Since the formalism presented here is valid for bosons as well as for fermions, an application to nuclear systems 
might be promising, provided that a good guess for $W(\rho)$ is found.
The effective theory point of view might 
be again of help as envisaged by results obtained recently in Ref.\cite{Rocco}. 
Work in that direction is in progress.  

{\bf Acknowledgement}

The authors would like to thanks L. Girlanda for useful discussions. 
Moreover, A.K. would like to thank early discussions on the subject with A. Polls and
G.O. would like to thank M. Calandra e S. Giorgini for useful discussions.


\begin{thebibliography}{10}
\bibitem{hohemberg1964} P. Hohenberg and W. Kohn, Phys. Rev. {\bf 136}, B864 (1964) 
\bibitem{kohn1965} W. Kohn and L.J. Sham, Phys. Rev. {\bf 140}, A1133 (1965)
\bibitem{vignale}G.F. Giuliani and G. Vignale, Quantum Theory of the Electron Liquid,
Cambridge University Press, UK, 2005
\bibitem{lipparini} E. Lipparini, Modern Many-Particle Physics, Atomic Gases, Quantum Dots and
Quantum Fluids, Wordl Scientific, Singapore, 2003
\bibitem{parr} R.G. Parr and W. Tangm Density-Functional Theory of Atoms and Molecules (International
Series of Monographs on Chemistry) (Oxford University Press, USA, 1994)
\bibitem{bender2003} M.B. Bender, P.-H. Heenen, and P.-G. Reinhard, Rev. Mod. Phys. {\bf 75}, 121 (2003)
\bibitem{schunck} N. Schunk, ed., Energy Density Functional Methods for Atomic Nuclei, 2053-2563
 (IOP Publishing, 2019)
 \bibitem{Engel2007}J. Engel, Phys. Rev. C {\bf 75}, 014306 (2007), nucl-th-0610043
 \bibitem{Giraudetal2008}B.G. Giraud, B.K. Jennings, B.R. Barrett, Phys. Rev. A {\bf 78}, 032507 (2008), 
 \bibitem{Barnea2007}N. Barnea, Phys. Rev. C 76, 067302 (2007). 
 \bibitem{Giraud2008}B.G. Giraud, Phys. Rev. C 77, 014311 (2008); Phys. Rev. C {\bf 78}, 014307 (2008).

 \bibitem{Messud2009}J. Messud, M. Bender, E. Suraud, Phys. Rev. C {\bf 80}, 054314 (2009). 

 \bibitem{Chamel2010}N. Chamel, Phys. Rev. C {\bf 82}, 061307(R) (2010). 

\bibitem{Duguet2010}T. Duguet, J. Sadoudi, J. Phys. G {\bf 37}, 064009 (2010)

 \bibitem{Messud2013}J. Messud, Phys. Rev. C {\bf 87}, 024302 (2013). [Addendum:
Phys. Rev.C87,no.2,029904(2013)].
\bibitem{marino} F. Marino, C. Barbieri, G. Colo', A. Lovato, F. Pederiva, X. Roca-Maza, and
 E. Vigezzi, arXiv:2103.14480[nucl-th]

\bibitem{Efros1972}V. D. Efros, Sov. J. Nucl. Phys. {\bf 15}, 128 (1972).
\bibitem{fabre1983}M. Fabre de la Ripelle, Ann. Phys. (N.Y.) {\bf 147}, 281 (1983)
\bibitem{Barnea2000}  N. Barnea, W. Leidemann, G. Orlandini, Phys.Rev. C {\bf 61} 054001 (2000) 054001 
\bibitem{gattobigio2009}M. Gattobigio, A. Kievsky, M. Viviani, and P. Barletta,
Phys. Rev. A {\bf 79}, 032513 (2009)
\bibitem{kievsky2008}
A. Kievsky, S. Rosati, M. Viviani, L.E. Marcucci and L. Girlanda, J. Phys. G {\bf 35}, 063101 (2008).

%\bibitem{smirnov}Y.F. Smirnov and K.V. Shitikova, Fiz. Elem. Chastits. At. Yadra. {\bf 8}, 847 (1977)
%[Sov. J. Part. Nucl. {\bf 8}, 344 (1977)]
%\bibitem{GP}V.L. Ginzburg and L.P. Pitaevskii, Zh. Eksp. Teor. Fiz. {\bf 34}, 1240 (1958)
%[Sov. Phys.JETP {\bf 7}, 858 (1958)]; E.P. Gross, J. Math. Phys. {\bf 4}, 195 (1963)
\bibitem{fabre1979} M. Fabre de la Ripelle and J. Navarro, Ann. Phys. (N.Y.) {\bf 123}, 185 (1979)
\bibitem{fabre1982} M. Fabre de la Ripelle, H. Fiedeldey, and G. Wiechers, Ann. Phys. {\bf 138}, 275 (1982)
\bibitem{greene1998}J.L. Bohn, B.D. Esry, and C.H. Greene, Phys. Rev. A {\bf 58}, 584 (1998)
%\bibitem{simonov} Yu.. A. Simonov, Sov. J. Nucl. Phys. {\bf 7}, 722 (1968)
\bibitem{aziz1}R.A. Aziz, V.P.S. Nain, J.S. Carley, W.L. Taylor, and G.T.  McConville, 
        J. Chem. Phys. {\bf 70}, 4330 (1979).
\bibitem{bira} H.W. Hammer, S. K{\"o}nig, and U.{van Kolck}, {Rev. Mod. Phys.} {\bf 92}, 025004 (2020)
\bibitem{bazak} B. Bazak, M. Eliyahu, and U. {van Kolck}, {Phys. Rev. A} {\bf 94}, 052502 (2016)
\bibitem{thomas} L.H. Thomas, {Phys. Rev.} {\bf 47}, 903 (1935)
\bibitem{timofeyuk1} N.K. Timofeyuk, Phys. Rev. A {\bf 86}, 032507 (2012)
\bibitem{timofeyuk2} N.K. Timofeyuk, Phys. Rev. A {\bf 91}, 042513 (2015)
\bibitem{pandha1} V. R. Pandharipande, J.G. Zabolitzky, S.C. Pieper, R.B.
Wiringa, and U. Helmbrecht,
        Phys. Rev. Lett. {\bf 50}, 1676 (1983).
\bibitem{dalfovo}F. Dalfovo, A. Lastri, L. Pricaupenko, S. Stringari, J. Treiner, Phys. Rev. B {\bf 52},
 1193 (1995)
\bibitem{artur1} A. Kievsky, A. Polls, B. Juli\'a-D\'\i az, and N.K. Timofeyuk,
         Phys. Rev A {\bf 96}, 040501(R) (2017).
\bibitem{artur2} A. Kievsky, A. Polls, B. Juli\'a-D\'\i az, N.K. Timofeyuk, and M. Gattobigio,
         Phys. Rev A {\bf 102}, 063320 (2020).
\bibitem{Rocco}R. Schiavilla, L. Girlanda, A. Gnech, A. Kievsky, A. Lovato, L.E. Marcucci, M. Piarulli,
and M. Viviani, Phys. Rev. C {\bf 103}, 054003 (2021)



\end{thebibliography}
\end{document}